%% file: draft.tex
\documentclass[conference]{IEEEtran}
\IEEEoverridecommandlockouts
\usepackage{algorithm,algorithmic,amsmath,amssymb,amsthm,bbm,cite,colortbl,comment,enumitem,graphicx,makecell,microtype,multirow,setspace,tabularx,url,xcolor}

\usepackage{hyperref}
\usepackage[utf8]{inputenc}
\usepackage[T1]{fontenc}
\usepackage[english]{babel}
\usepackage[figurename=Fig., labelsep=period, font=small]{caption}

\usepackage[nolist]{acronym}

\usepackage{tikz,pgfplots}
\usepgfplotslibrary{colormaps}
\usetikzlibrary{arrows, arrows.meta, backgrounds, calc, intersections, decorations.markings, decorations.pathreplacing, positioning, shapes, external}
\pgfplotsset{compat=newest}
\pgfplotsset{
/pgfplots/ybar legend/.style={
/pgfplots/legend image code/.code={%
\draw[##1,/tikz/.cd,bar width=0.1cm,yshift=-0.2em,bar shift=0.5*\pgfplotbarwidth]
plot coordinates {(0.5*\pgfplotbarwidth,0.6em) (2.5*\pgfplotbarwidth,0.4em) (4.5*\pgfplotbarwidth,0.2em)};},}}

\usepackage{titlesec}
\let\subparagraph\relax
\titlespacing{\section}{0pt}{5pt plus 2pt minus 1pt}{3pt plus 1pt minus 1pt} 
\titlespacing{\subsection}{0pt}{4pt plus 2pt minus 1pt}{2pt plus 1pt minus 1pt} 
\input{./Acronyms.tex}

\input{./Definitions.tex}

\title{Structured Wavenumber-Domain Channel Estimation for LoS XL-MIMO}
\author{
\IEEEauthorblockN{Nirosh Marthinu,$^{1}$
Prathapasinghe Dharmawansa,$^{2}$ and
Italo Atzeni$^{2}$}
\IEEEauthorblockA{$^{1}$ {Dept. of Computer Science, Electrical and Space Engineering, Lule\aa{} University of Technology, Sweden} \\
$^{2}$ Centre for Wireless Communications, University of Oulu, Finland \\
E-mails: nirosh.marthinu@ltu.se, \{prathapasinghe.kaluwadevage, italo.atzeni\}@oulu.fi}
\thanks{This work was supported by the Research Council of Finland (336449 Profi6, 348396 HIGH-6G, and 369116 6G~Flagship). Nirosh Marthinu was with the University of Oulu when this work was prepared.}}

\begin{document}

\maketitle

\begin{abstract}
Near-field line-of-sight (LoS) channels in extremely large-scale multiple-input multiple-output (XL-MIMO) systems exhibit distinctive structural properties in the wavenumber domain. In this paper, we develop a structured wavenumber-domain channel estimation framework that first obtains an initial channel estimate using classical estimators and then refines it through a windowing-based refinement step that suppresses out-of-band noise by exploiting the localized spatial-frequency power distribution. The resulting estimate is subsequently enhanced through a convex constrained formulation that exploits the sparsity and low-rank structure of the wavenumber-domain channel. The corresponding sparsity and rank constraint bounds are selected using a conservative worst-case characterization over the relative position of the transmitter and receiver. Numerical results show that the proposed framework improves estimation accuracy, particularly at low signal-to-noise ratio, by effectively leveraging the structural properties of the wavenumber-domain channel.
\end{abstract}


\acresetall

\section{Introduction}

\Ac{xlmimo} is emerging as a promising physical-layer technology for \ac{6G} wireless systems~\cite{bjornson2019_whatisnext}. Compared with conventional massive \ac{MIMO}, \acs{xlmimo} employs substantially larger array apertures and increasingly operates at frequencies spanning the FR3 to sub-THz ranges. Consequently, the transition to \acs{xlmimo} brings near-field propagation into practical relevance, in contrast to the predominantly far-field operation of conventional massive \ac{MIMO}. In this regime, the conventional planar-wave assumption becomes inaccurate and spherical wavefronts must be accounted for in channel modeling~\cite{cui2022_farfieldornearfield,zhang20236g}. Moreover, the reduced diffraction and scattering at such high frequencies make \ac{LoS} propagation increasingly dominant over \ac{NLoS} components~\cite{italo2025hexa}, rendering near-field \ac{LoS} channels particularly relevant for next-generation systems.

Accurate channel estimation is essential for realizing the full potential of \ac{LoS} \acs{xlmimo} systems. However, as the number of antennas grows, the channel dimension increases substantially, making conventional estimation approaches computationally demanding unless the underlying channel structure is explicitly exploited~\cite{bacci2024_mmse_channel}. In this context, the Fourier plane-wave series expansion provides a physically motivated representation of the near-field channel by approximating a spherical wave by a finite sum of plane waves~\cite{pizzo2022_holographicmimo}. This transformation maps the dense antenna-domain channel into a sparse wavenumber-domain representation~\cite{Wang2024XL_MIMO}. Beyond sparsity, near-field \ac{LoS} channels exhibit a low-rank structure, as the number of spatial degrees of freedom is fundamentally constrained by the array geometry and propagation distance~\cite{pizzo2022_holographicmimo}. Several recent works have jointly exploited these structural properties for channel estimation~\cite{guo2025_geometricprior,tang2024_jointvisibility}. However, the localized power distribution of the wavenumber-domain channel has not been explicitly leveraged to systematically determine the sparsity and rank constraint bounds.

In this paper, we develop a structured wavenumber-domain channel estimation framework for \ac{LoS} \acs{xlmimo}. The framework first obtains an initial channel estimate using classical estimators and then refines it through a \ac{WBR} step that leverages the localized spatial-frequency power distribution to suppress out-of-band noise. The resulting estimate is subsequently enhanced through a convex constrained formulation that exploits the inherent sparsity and low-rank structure of the wavenumber-domain channel, with the sparsity and rank constraint bounds selected using a conservative worst-case characterization over the relative position of the \ac{Tx} and \ac{Rx}. Numerical results evaluate the proposed framework and demonstrate the gains achieved by exploiting the structural properties of the wavenumber-domain channel, particularly at low \ac{SNR}.

\section{System Model}

In this section, we introduce the considered \ac{LoS} \acs{xlmimo} system model and derive the corresponding wavenumber-domain representation. We consider a narrowband communication with operating frequency $f$ and resulting wavelength $\lambda = \frac{c}{f}$, where $c$ is the speed of light in vacuum. The corresponding wavenumber is denoted by $k_0 = \frac{2\pi}{\lambda}$.

\subsection{LoS XL-MIMO System Model}

\begin{figure}
\centering
\input{Figures/fig_1.tex}
\vspace{1mm}
\caption{Considered point-to-point \ac{LoS} \acs{xlmimo} system.}
\label{system_model}
\end{figure}
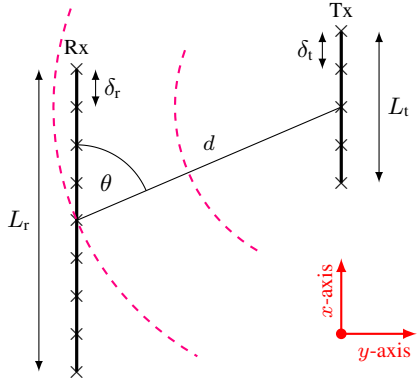

Consider the point-to-point \ac{LoS} \acs{xlmimo} system illustrated in Fig.~\ref{system_model}, where the \ac{Rx} is a base station equipped with an extremely large antenna array, and both the \ac{Tx} and \ac{Rx} employ \acp{ULA}. Let $N_{\textrm{r}}$ and $N_{\textrm{t}}$ denote the numbers of antennas at the \ac{Rx} and \ac{Tx}, respectively. The corresponding array lengths are given by $L_{\textrm{r}}=\delta_{\textrm{r}}(N_{\textrm{r}}-1)\approx\delta_{\textrm{r}}N_{\textrm{r}}$ and $L_{\textrm{t}}=\delta_{\textrm{t}}(N_{\textrm{t}}-1)$, where $\delta_{\textrm{r}}$ and $\delta_{\textrm{t}}$ denote the antenna spacings at the \ac{Rx} and \ac{Tx}, respectively.\footnote{Since the \ac{Rx} is equipped with an extremely large antenna array with $N_{\textrm{r}}\gg1$, the approximation $N_{\textrm{r}}-1\approx N_{\textrm{r}}$ is reasonable.} The antenna-domain channel matrix is denoted by $\mathbf{H}\in\mathbb{C}^{N_{\textrm{r}}\times N_{\textrm{t}}}$.

We assume that the two \acp{ULA} are parallel to the $x$-axis. Let $d$ denote the distance between the array centers and let $\theta \in (0, \pi)$ represent the angle between the positive $x$-axis and the line connecting the array centers. The $m$-th \ac{Rx} antenna is located at $\mathbf{r}_m = [r_{m,x},r_{m,y}]^{\mathrm{T}} = \big[(m - \frac{N_{\textrm{r}}+1}{2})\delta_{\textrm{r}}, 0\big]^{\mathrm{T}} \in \mathbb{R}^2$, for $m = 1, \ldots, N_{\textrm{r}}$, whereas the $n$-th \ac{Tx} antenna is located at $\mathbf{t}_n = [t_{n,x},t_{n,y}]^{\mathrm{T}}= \big[d\cos\theta + (n - \frac{N_{\textrm{t}}+1}{2})\delta_{\textrm{t}},d\sin\theta\big]^{\mathrm{T}} \in \mathbb{R}^2$, for $n = 1, \ldots, N_{\textrm{t}}$. Then, the $(m,n)$-th channel coefficient is given by~\cite{Wang2024XL_MIMO}
\begin{align}
    [\mathbf{H}]_{m,n} = \frac{\lambda}{4\pi}\frac{\exp(\mathrm{j}k_0 d_{m,n})}{d_{m,n}},
\label{perTXperRXantennachannel}
\end{align}
where $\mathrm{j}=\sqrt{-1}$ denotes the imaginary unit and $d_{m,n} = \|\mathbf{r}_m - \mathbf{t}_n\|_2$ is the distance between the $m$-th receive antenna and the $n$-th transmit antenna. As shown in \eqref{perTXperRXantennachannel}, the near-field \ac{LoS} channel is uniquely determined by the operating wavelength and the relative positions of the \ac{Tx} and \ac{Rx} antennas.

\subsection{Wavenumber-Domain Channel Representation}
\label{WND_representation}

To facilitate the analysis and estimation of the near-field \ac{LoS} channel, we approximate the spherical wave in \eqref{perTXperRXantennachannel} by a finite sum of plane waves using the Fourier plane-wave series expansion~\cite{pizzo2022_holographicmimo,Wang2024XL_MIMO}, where the approximation becomes asymptotically exact for $\frac{L_{\textrm{r}}}{\lambda} \to \infty$. Let
\begin{align}
\mathcal{S} = \left\{ \zeta \in \mathbb{Z} : -k_0 \leq \Delta \zeta < k_0 \right\}
\label{eq:spacial_frequiencies}
\end{align}
denote the set of spatial-frequency indices associated with the retained plane waves, where $\Delta = \frac{2\pi}{L_{\textrm{r}}}$ represents the spatial-frequency sampling interval and $L=|\mathcal{S}|$ is the number of discrete spatial-frequency components. Accordingly, the antenna-domain channel in \eqref{perTXperRXantennachannel} can be approximated as
\begin{align}
    [\mathbf{H}]_{m,n} \approx \sum_{\ell=1}^{L} g_{\ell,n} \big[\mathbf{a}(\zeta_\ell)\big]_m,
    \label{eq:WNDapproximation}
\end{align}
where, for the $\ell$-th spatial-frequency component, $g_{\ell,n} \in \Compl$ denotes the complex gain corresponding to the $n$-th \ac{Tx} antenna and $\mathbf{a}(\zeta_\ell)\in \mathbb{C}^{N_{\textrm{r}}\times 1}$ is the array response vector, with $m$-th element given by $\big[\mathbf{a}(\zeta_\ell)\big]_m = \frac{1}{\sqrt{N_{\textrm{r}}}} \exp(\mathrm{j} \Delta \zeta_\ell r_{m,x})$. Then, the wavenumber-domain codebook $\mathbf{F}_{\textrm{w}} = \big[\mathbf{a}(\zeta_1), \ldots,\mathbf{a}(\zeta_L) \big] \in \mathbb{C}^{N_{\textrm{r}}\times L}$ is constructed by stacking the $L$ array response vectors. Hence, \eqref{eq:WNDapproximation} can be expressed in matrix form as $\mathbf{H} \approx \mathbf{F}_{\textrm{w}}\mathbf{H}_{\textrm{w}}$, where $\mathbf{H}_{\textrm{w}}\in\mathbb{C}^{L \times N_{\textrm{t}}}$ is the wavenumber-domain channel matrix, with $(\ell,n)$-th element given by $[\mathbf{H}_{\textrm{w}}]_{\ell,n} = g_{\ell,n}$.

In this work, we consider $\delta_\textrm{r} = \frac{\lambda}{2}$, which reduces \eqref{eq:spacial_frequiencies} to $\mathcal{S} = \big\{ \zeta \in \mathbb{Z} : -\frac{N_{\textrm{r}}-1}{2} \leq \zeta < \frac{N_{\textrm{r}}-1}{2} \big\}$, yielding $L = N_\textrm{r} - 1 \approx N_\textrm{r}$. In this regime, the columns of $\mathbf{F}_{\textrm{w}}$ become asymptotically orthonormal for $N_{\textrm{r}} \to \infty$, satisfying $\mathbf{F}_{\textrm{w}}^{\mathrm H} \mathbf{F}_{\textrm{w}} \approx \mathbf{I}_{L}$. Therefore, the wavenumber-domain channel matrix can be obtained as
\begin{align}
\mathbf{H}_{\textrm{w}} \approx \mathbf{F}_{\textrm{w}}^{\mathrm H} \mathbf{H}.
\label{eq:WNDapproximationMat}
\end{align}

\section{Sparse and Low-Rank Channel Estimation}

Due to the finite array aperture, only a limited subset of the spatial-frequency components in $\mathcal{S}$ contributes significantly to the wavenumber-domain channel representation in \eqref{eq:WNDapproximation}~\cite{Wang2024XL_MIMO}. These dominant components are typically concentrated over a localized spatial-frequency region whose location depends on the relative \ac{Tx}--\ac{Rx} position, giving rise to a sparse wavenumber-domain channel representation. Moreover, the channel responses associated with different \ac{Tx} antennas are governed by the same propagation geometry and therefore span only a limited number of spatial degrees of freedom, resulting in a low-rank structure. In this section, we jointly exploit these two properties to improve the channel estimation accuracy.

Assume that an initial channel estimate $\tilde{\mathbf{H}}_{\textrm{w}} \in \Compl^{L \times N_{\textrm{t}}}$ is available. This can be obtained, e.g., via \ac{LS} estimation or other low-complexity methods, as described in Section~\ref{initial_estimation}. The estimate can be improved by seeking a channel matrix that remains close to $\tilde{\mathbf{H}}_{\textrm{w}}$ while satisfying sparsity and rank constraints. Accordingly, an enhanced channel estimate $\hat{\mathbf{H}}_{\textrm{w}} \in \Compl^{L \times N_{\textrm{t}}}$ can be obtained by solving
\begin{align}
\begin{array}{cl}
\displaystyle \underset{\hat{\mathbf{H}}_{\textrm{w}}}{\mathrm{minimize}} & \|\hat{\mathbf{H}}_{\textrm{w}}-\tilde{\mathbf{H}}_{\textrm{w}}\|_\mathrm{F}^2\\
\mathrm{subject~to} & \big\|\mathrm{vec}(\hat{\mathbf{H}}_{\textrm{w}})\big\|_0 \leq \bar\alpha, \\
& \mathrm{rank}(\hat{\mathbf{H}}_{\textrm{w}}) \leq \bar\beta,
\end{array}
\label{eq:nonconvex problem}
\end{align}
where $\|\cdot\|_\mathrm{F}$ is the Frobenius norm, $\|\cdot\|_0$ is the $\ell_{0}$-norm indicating the number of nonzero elements, $\mathrm{vec}(\cdot)$ is the vectorization operator, and $\mathrm{rank}(\cdot)$ is the rank operator. The parameters $\bar\alpha,\bar\beta \in \mathbb{Z}_+$ denote the sparsity and rank constraint bounds, respectively.

The problem in \eqref{eq:nonconvex problem} is non-convex and NP-hard due to the combinatorial nature of both the sparsity and rank constraints. Adopting standard convex relaxation approaches~\cite{boyd2004convex,candes_lowrankmatrixcompletion}, we replace the $\ell_0$-norm by the $\ell_1$-norm and the rank operator by the nuclear norm, yielding the approximate problem
\begin{align}
\begin{array}{cl}
\displaystyle \underset{\hat{\mathbf{H}}_{\textrm{w}}}{\mathrm{minimize}} & \|\hat{\mathbf{H}}_{\textrm{w}}-\tilde{\mathbf{H}}_{\textrm{w}}\|_\mathrm{F}^2\\
\mathrm{subject~to} & \big\|\mathrm{vec}(\hat{\mathbf{H}}_{\textrm{w}})\big\|_1 \leq \alpha, \\
& \|\hat{\mathbf{H}}_{\textrm{w}}\|_{*} \leq \beta,
\end{array}
\label{eq:convex_problem}
\end{align}
where $\|\cdot\|_1$ and $\|\cdot\|_{*}$ are the $\ell_1$-norm and nuclear norm, respectively. The parameters $\alpha, \beta \in \mathbb{R}_+$ denote the $\ell_1$-norm and nuclear-norm constraint bounds, respectively, which control the strength of the corresponding relaxed sparsity and rank regularizations. The problem in \eqref{eq:convex_problem} is convex and can be efficiently solved via standard solvers such as CVX~\cite{cvx}.

\section{Constraint-Bound Selection, Initial Channel Estimation, and WBR}

The problem in \eqref{eq:convex_problem} requires two key inputs: the initial channel estimate $\tilde{\mathbf{H}}_{\textrm{w}}$ and the constraint bounds $\alpha$ and $\beta$. In this section, we describe how these inputs are obtained. First, we determine $\alpha$ and $\beta$ through a conservative worst-case approach. Then, we discuss three methods for obtaining $\tilde{\mathbf{H}}_{\textrm{w}}$, namely, \ac{LS}~\cite{kay1993_estimation}, \ac{LASSO}~\cite{tibshirani1996_lasso}, and \ac{IRL1}~\cite{candes2008_reweighted}. Finally, we propose an intermediate post-processing step, referred to as \textit{\acf{WBR}}, to refine the initial channel estimate before it is used in \eqref{eq:convex_problem}.

\subsection{Constraint-Bound Selection}
\label{constraint_bounds}

The wavenumber-domain channel structure depends on the relative position of the \ac{Tx} and \ac{Rx} \acp{ULA}, making it challenging to select universal values for $\alpha$ and $\beta$ in \eqref{eq:convex_problem}. To address this, we investigate how the $\ell_1$-norm and nuclear norm of $\mathbf{H}_{\textrm{w}}$ vary with the numbers of \ac{Tx} and \ac{Rx} antennas $N_{\textrm{t}}$ and $N_{\textrm{r}}$ as well as the distance $d$ and angle $\theta$ between the \ac{Tx} and \ac{Rx}.

Since the exact \ac{Tx} array position is unknown in practice, we adopt a conservative worst-case approach and select $\alpha$ and $\beta$ as upper bounds over a considered distance-angle region $\mathcal{G}$. For each fixed $N_{\textrm{r}}$, these bounds are defined as
\begin{align}
\alpha(N_{\textrm{t}}) &= \max_{(\theta,d)\in\mathcal{G}} \big\| \mathrm{vec}\big(\mathbf{H}_{\textrm{w}}(\theta,d)\big) \big\|_1,
\label{eq:alpha_def}\\
\beta(N_{\textrm{t}}) &= \max_{(\theta,d)\in\mathcal{G}} \big\| \mathbf{H}_{\textrm{w}}(\theta,d) \big\|_*.
\label{eq:beta_def}
\end{align}
This choice ensures that the constraints in \eqref{eq:convex_problem} accommodate all possible \ac{Tx} array placements. If prior information on $\theta$ or $d$ is available, the selection can be restricted to the corresponding uncertainty region, yielding less conservative bounds.

\begin{figure}
\centering
\input{Figures/fig_2}
\vspace{-3mm}
\caption{Constraint bounds $\alpha$ and $\beta$ versus number of transmit antennas $N_{\textrm{t}}$. Left: $\alpha$ as a function of $\sqrt{N_{\textrm{t}}}$. Right: $\beta$ as a function of $N_{\textrm{t}}$.}
\label{fig:alphabeeta_with_Nt}
\end{figure}
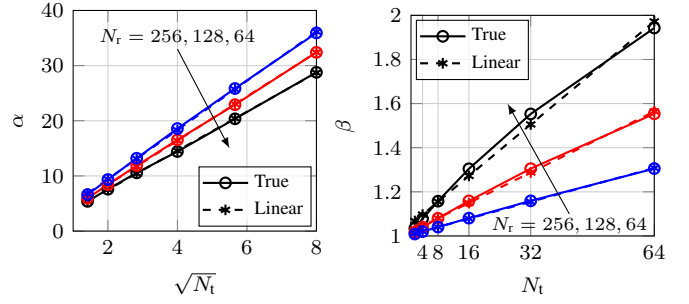

As a representative example, we consider $\mathcal{G}=\big\{(\theta,d):\theta\in\big(0,\frac{\pi}{2}\big],\, d\in[0.15\,d_{\textrm{F}},d_{\textrm{F}}]\big\}$, where $d_{\textrm{F}} = 2\frac{L_{\textrm{r}}^2}{\lambda}$ denotes the Fraunhofer distance with respect to the \ac{Rx} array. Due to the symmetry of the array geometry, the range $\theta\in \big(\frac{\pi}{2},\pi\big]$ yields an equivalent behavior to $\theta\in\big[0,\frac{\pi}{2}\big)$; hence, considering only $\theta\in\big(0,\frac{\pi}{2}\big]$ is sufficient to determine the constraint bounds. The scaling behavior is evaluated for $N_{\textrm{r}}\in\{64,128,256\}$ and $N_{\textrm{t}}\in\{2,4,8,16,32,64\}$. For each $(N_{\textrm{t}},N_{\textrm{r}})$ combination, the corresponding values of $\alpha$ and $\beta$, satisfying \eqref{eq:alpha_def} and \eqref{eq:beta_def}, respectively, are obtained through an exhaustive search over $\mathcal{G}$. Fig.~\ref{fig:alphabeeta_with_Nt} shows the resulting values of $\alpha$ and $\beta$ as functions of $N_{\textrm{t}}$ for different values of $N_{\textrm{r}}$. For a fixed $N_{\textrm{r}}$, $\alpha$ scales approximately with $\sqrt{N_{\textrm{t}}}$, while $\beta$ scales approximately with $N_{\textrm{t}}$. These trends are modeled as
\begin{align}
\alpha \approx \mathrm{m_1}\sqrt{N_{\textrm{t}}} + \mathrm{c_1}, \quad
\beta  \approx \mathrm{m_2} N_{\textrm{t}} + \mathrm{c_2}, \label{eq:alp_beta_eqs}
\end{align}
where the coefficients $\mathrm{m_1}$, $\mathrm{m_2}$, $\mathrm{c_1}$, and $\mathrm{c_2}$ depend on $N_\textrm{r}$ and can be found via curve fitting.

\subsection{Initial Channel Estimation}
\label{initial_estimation}

After fixing the constraint bounds $\alpha$ and $\beta$, the remaining input to \eqref{eq:convex_problem} is the initial channel estimate $\tilde{\mathbf{H}}_{\textrm{w}}$. We consider three well-established methods for obtaining this estimate, each offering a different trade-off between computational complexity and estimation accuracy. Then, in Section~\ref{sec:windowing}, we present the proposed \ac{WBR}, which can be applied as a post-processing step to any of these initial channel estimates.

Let $\mathbf{X}_{\textrm{p}} \in \mathbb{C}^{N_{\textrm{t}} \times \tau}$ denote the pilot matrix transmitted by the \ac{Tx} for channel estimation, where $\tau \geq N_{\textrm{t}}$ denotes the pilot length. The rows of $\mathbf{X}_{\textrm{p}}$ are assumed to be orthogonal, i.e., $\mathbf{X}_{\textrm{p}}\mathbf{X}_{\textrm{p}}^\mathrm{H} = \tau\mathbf{I}_{N_{\textrm{t}}}$. The received pilot signal at the \ac{Rx} is given by
\begin{align}
    \mathbf{Y} = \sqrt{\rho}\mathbf{H}\mathbf{X}_{\textrm{p}} + \mathbf{Z} \in \mathbb{C}^{N_{\textrm{r}} \times \tau},
    \label{eq:estimation_model}
\end{align}
where $\mathbf{Z} \in \mathbb{C}^{N_{\textrm{r}} \times \tau}$ is a matrix of \ac{AWGN} with \ac{iid} $\mathcal{CN}(0, 1)$ elements and $\rho$ denotes the transmit \ac{SNR}. Using the transformation in \eqref{eq:WNDapproximationMat}, the received pilot signal in \eqref{eq:estimation_model} can be expressed in the wavenumber domain as
\begin{align}
    \mathbf{Y}_{\textrm{w}} &=\mathbf{F}_{\textrm{w}}^\mathrm{H}\mathbf{Y} \approx \sqrt{\rho}\mathbf{H}_{\textrm{w}}\mathbf{X}_{\textrm{p}} + \mathbf{Z}_{\textrm{w}} \in \mathbb{C}^{L \times \tau},
    \label{eq:WND estimation_model}
\end{align}
where $\mathbf{Z}_{\textrm{w}}=\mathbf{F}_{\textrm{w}}^\mathrm{H}\mathbf{Z} \in \mathbb{C}^{L \times \tau}$ is the transformed \ac{AWGN} matrix with approximately \ac{iid} $\mathcal{CN}(0,1)$ elements.

Now, define $\mathbf{y}_{\textrm{w}} = \mathrm{vec}(\mathbf{Y}_{\textrm{w}}) \in \mathbb{C}^{L\tau\times 1}$ and $\mathbf{\overline{X}}_{\textrm{p}} = \mathbf{X}_{\textrm{p}}^\mathrm{T} \otimes \mathbf{I}_{L} \in \mathbb{C}^{L \tau \times LN_{\textrm{t}}}$, where $\otimes$ denotes the Kronecker product. Based on the observation $\mathbf{y}_{\textrm{w}}$, we obtain an estimate of the vectorized channel $\tilde{\mathbf{h}}_{\textrm{w}} = \vec (\tilde{\mathbf{H}}_{\textrm{w}}) \in \mathbb{C}^{L N_{\textrm{t}} \times 1}$ using \ac{LS}, \ac{LASSO}, or \ac{IRL1}, and then reshape it into matrix form as $\tilde{\mathbf{H}}_{\textrm{w}} = \mathrm{unvec}(\tilde{\mathbf{h}}_{\textrm{w}})$.

\textit{\textbf{\ac{LS} estimator.}} Assuming that the rows of $\mathbf{X}_{\mathrm{p}}$ are orthogonal, the \ac{LS} estimate is obtained as
\begin{align}
    \tilde{\mathbf{h}}_{\textrm{w,LS}}
&= \frac{1}{\sqrt{\rho} \tau} \mathbf{\overline{X}}_{\textrm{p}}^\mathrm{H} \mathbf{y}_{\textrm{w}}.
\label{eq:ls_estimation}
\end{align}

\textit{\textbf{\ac{LASSO} estimator.}} Since the wavenumber-domain channel is sparse, the \ac{LASSO} estimator can be used to promote sparsity via $\ell_1$-norm regularization \cite{tibshirani1996_lasso}. Its estimate is given by
\begin{align}
    \tilde{\mathbf{h}}_{\text{w,LASSO}} = \underset{\tilde{\mathbf{h}}_{\textrm{w}}}{\argmin} \,
    \|\mathbf{y}_{\textrm{w}} - \sqrt{\rho}\mathbf{\overline{X}}_{\textrm{p}}\tilde{\mathbf{h}}_{\textrm{w}}\|_2^2 + \nu \|\tilde{\mathbf{h}}_{\textrm{w}}\|_1,
    \label{eq:lasso_optimization}
\end{align}
where $\nu > 0$ controls the trade-off between data fidelity and sparsity.

\textit{\textbf{\ac{IRL1} estimator.}} \ac{IRL1} iteratively solves a sequence of weighted $\ell_1$-norm problems, heavily penalizing small coefficients while preserving large ones \cite{candes2008_reweighted}. The channel estimate at iteration~$k$ is given by
\begin{align}
\begin{array}{rcl}
\tilde{\mathbf{h}}_{\textrm{w,IRL1}}^{(k)} = \! \! \! \! & \displaystyle \underset{\tilde{\mathbf{h}}_{\textrm{w}}}{\argmin} & \displaystyle \sum_{i=1}^{LN_{\textrm{t}}} w_i^{(k-1)} \big|[\tilde{\mathbf{h}}_{\textrm{w}}]_i\big|\\
& \mathrm{s.t.} & \|\mathbf{y}_{\textrm{w}} - \sqrt{\rho} \mathbf{\overline{X}}_{\textrm{p}}\tilde{\mathbf{h}}_{\textrm{w}}\|_2 \leq \varphi,
\end{array}
\label{eq:irl1_problem}
\end{align}
where $\varphi > 0$ is chosen such that the true channel vector $\mathbf{h}_{\textrm{w}} = \mathrm{vec}(\mathbf{H}_{\textrm{w}})\in \mathbb{C}^{LN_{\textrm{t}} \times 1}$ is feasible for \eqref{eq:irl1_problem}. Each weight is initialized as $w_i^{(0)}=1$ and updated as $w_i^{(k)} = \big(\eta + |[\mathbf{h}^{(k)}]_i|\big)^{-1}$, where $\eta \ll 1$ ensures numerical stability.

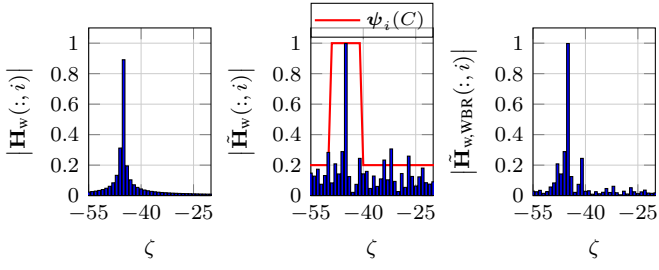
\begin{figure}
\centering
\input{Figures/fig_3}
\vspace{-3mm}
\caption{Effect of \ac{WBR} on the magnitude of the initial wavenumber-domain channel estimate, with $\gamma = 0.2$. Left: true channel $\big|\mathbf{H}_\textrm{w}(:,i)\big|$. Middle: initial channel estimate $\big|\tilde{\mathbf{H}}_\textrm{w}(:,i)\big|$ overlaid with windowing vector $\boldsymbol{\psi}_i(C)$. Right: \ac{WBR}-refined channel estimate $\big|\tilde{\mathbf{H}}_\textrm{w,WBR}(:,i)\big|$.}
\label{fig:effect_of_WBR}
\end{figure}

\subsection{Windowing-Based Refinement}
\label{sec:windowing}

The above estimators do not explicitly exploit the localized power distribution across spatial-frequency components characterized in Section~\ref{constraint_bounds}. Their result can therefore be further refined before being used in \eqref{eq:convex_problem}. To this end, we propose applying \ac{WBR}, described next, to the initial channel estimate $\tilde{\mathbf{H}}_{\textrm{w}}$ to suppress out-of-band noise while preserving the dominant spatial-frequency components.

Given the initial channel estimate, let $\zeta_{q_i}$ denote the index of the dominant spatial-frequency component of the $i$-th \ac{Tx} antenna, where $q_i$ is determined as
\begin{align}
q_i = \underset{\ell = 1,\ldots,L}{\argmax}\; \big|[\tilde{\mathbf{H}}_{\textrm{w}}]_{\ell,i}\big|.
\label{eq:mu_identification}
\end{align}
Considering an odd window length $C$ (without loss of generality), let $\boldsymbol{\psi}_i(C) \in\mathbb{R}_+^{L}$ denote the windowing vector for the $i$-th \ac{Tx} antenna, with elements
\begin{align}
   \hspace{-2mm} \big[\boldsymbol{\psi}_{i}(C)\big]_{l} =
    \begin{cases}
        1 & \text{for } |\zeta_\ell - \zeta_{q_i}| \leq \big\lfloor \frac{C}{2} \big\rfloor,\\
        \gamma & \text{otherwise},
    \end{cases}
    \quad \ell=1,\ldots,L,
    \label{eq:windowing_function}
\end{align}
where $\gamma\in[0,1)$ controls the suppression applied to the out-of-band spatial-frequency components. In particular, $\gamma=0$ corresponds to hard thresholding, whereas $\gamma>0$ retains residual contributions from out-of-band components. The windowing matrix is defined as $\mathbf{\Psi}(C)=\big[\boldsymbol{\psi}_1(C),\ldots,\boldsymbol{\psi}_{N_{\textrm{t}}}(C)\big]\in\mathbb{R}^{L\times N_{\textrm{t}}}$, and the refined channel estimate obtained after applying \ac{WBR} is given by
\begin{align}
\tilde{\mathbf{H}}_{\textrm{w,WBR}}(C) = \tilde{\mathbf{H}}_{\textrm{w}} \odot \mathbf{\Psi}(C),
\label{eq:shrinkage_estimate}
\end{align}
where $\odot$ denotes the element-wise product. The impact of the proposed \ac{WBR} on the initial channel estimation is shown in Fig.~\ref{fig:effect_of_WBR} for $\gamma=0.2$.

Ideally, the optimal window length $C^\star$, which determines the number of spatial-frequency components retained at full weight, could be selected as
\begin{align}
C^\star = \underset{C \in \mathbb{Z}_+}{\argmin} \, \big\|\mathbf{H}_{\textrm{w}} - \tilde{\mathbf{H}}_{\textrm{w,WBR}}(C)\big\|_\mathrm{F}^2.
\label{eq:optimal_C}
\end{align}
In practice, however, the true channel $\mathbf{H}_{\textrm{w}}$ is unknown, making \eqref{eq:optimal_C} infeasible. Therefore, we determine a suboptimal window length, denoted by $C_{\textrm{sub}}^\star$, empirically by analyzing the power distribution of the true wavenumber-domain channel. For the $i$-th \ac{Tx} antenna, let $\mathcal{E}_i=\big\{\big|[\mathbf{H}_{\textrm{w}}]_{\ell,i}\big|^2\big\}_{\ell=1}^{L}$ denote the set of spatial-frequency power values and let $e_i^{(j)}$ represent the $j$-th largest component in $\mathcal{E}_i$. The minimum window length $C_i$ required to capture at least a fraction $1 - \epsilon$ of the total power is then defined as the smallest positive integer satisfying
\begin{align}
\frac{\sum_{j=1}^{C_i} e_i^{(j)}}{\sum_{\ell=1}^{L} \big|[\mathbf{H}_{\textrm{w}}]_{\ell,i}\big|^2} \geq 1 - \epsilon,
\label{eq:power_threshold}
\end{align}
where $\epsilon \in (0, 1)$ is a tunable power tolerance. To obtain a single window length applicable to all \ac{Tx} antennas, we select
\begin{align}
C_{\textrm{sub}}^\star =  \max_{i=1,\ldots,N_{\textrm{t}}}C_i.
\label{eq:system_C}
\end{align}

\begin{figure}
\centering
\input{Figures/fig_4}
\vspace{-3mm}
\caption{Effective number of spatial-frequency components $N_\text{eff}$ as a function of the angle~$\theta$ and distance~$d$ between the \ac{Tx} and \ac{Rx}. The 3D plot illustrates the joint dependence on angle and distance, while the subplots show slices at fixed distances.}
\label{fig:analysis_C}
\end{figure}
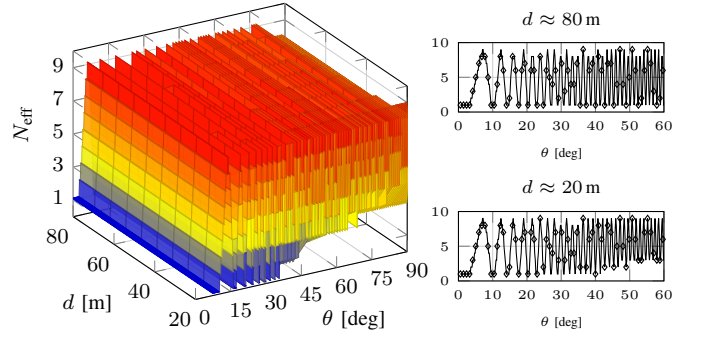

As a representative example, we determine a fixed value of $C_{\textrm{sub}}^\star$ that is suitable for all considered \ac{Tx} positions by evaluating \eqref{eq:power_threshold} and \eqref{eq:system_C} over $\theta\in\big(0,\frac{\pi}{2}\big]$ and $d\in[0.15d_{\textrm{F}},d_{\textrm{F}}]$, with $\epsilon=0.05$. Fig.~\ref{fig:analysis_C} shows the effective number of spatial-frequency components $N_\textrm{eff}$ as a function of the angle and distance between the \ac{Tx} and \ac{Rx}. Across all considered configurations, retaining $9$ spatial-frequency components is sufficient to capture a fraction $1-\epsilon$ of the total power.~Accordingly, we set $C_{\textrm{sub}}^\star=9$ in the subsequent numerical results.

\section{Numerical Results}
\label{sec:results}

In this section, we evaluate the performance of the proposed channel estimation framework, which first obtains an initial channel estimate using the methods described in Section~\ref{initial_estimation}, then refines it via \ac{WBR} presented in Section~\ref{sec:windowing}, and finally solves \eqref{eq:convex_problem} using the constraint bounds selected according to Section~\ref{constraint_bounds}. We normalize the channel as $\| \mathbf{H} \|_{\mathrm{F}}^{2} = 1$ and measure the estimation performance using the per-coefficient \ac{MSE} defined as $\textrm{MSE} = \frac{1}{N_{\textrm{t}} N_{\textrm{r}}} \Exp \big[ \| \mathbf{H} - \hat{\mathbf{H}} \|_{\mathrm{F}}^{2} \big]$. The simulations consider $\rho\in[0,30]$~dB and the results are averaged over $100$ independent channel and noise realizations for each value of $\rho$. The \ac{Tx} and \ac{Rx} employ $N_{\textrm{t}}=8$ and $N_{\textrm{r}}=256$ antennas, respectively, with the pilot length set to $\tau = N_{\textrm{t}}$. The operating frequency is set to $f=30$~GHz, with antenna spacings $\delta_{\textrm{t}}=\delta_{\textrm{r}}=\frac{\lambda}{2}$. For this configuration, \eqref{eq:alpha_def} and \eqref{eq:beta_def} yield $\alpha=13.1145$ and $\beta=1.0404$, respectively, corresponding to $\mathrm{m}_1=4.4528$, $\mathrm{c}_1=0.5201$, $\mathrm{m}_2=0.0048$, and $\mathrm{c}_2=1.0020$ in \eqref{eq:alp_beta_eqs}. For each realization, the angle and distance between the \ac{Tx} and \ac{Rx} are drawn as $\theta\sim\mathcal{U}(0,\pi)$ and $d\sim\mathcal{U}(0.15d_{\textrm{F}},0.9d_{\textrm{F}})$, respectively. Lastly, $\gamma=0.2$ is assumed for \ac{WBR}.

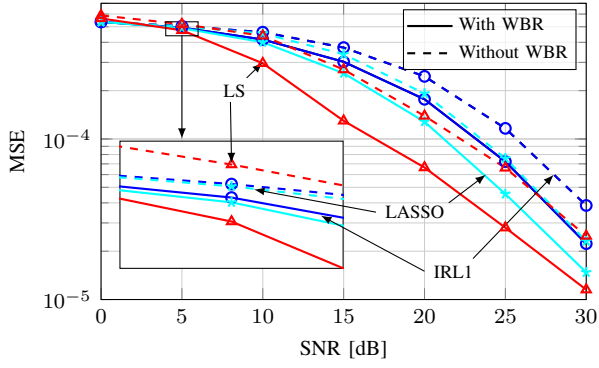
\begin{figure}
\centering
\input{Figures/fig_5}
\caption{\ac{MSE} versus \ac{SNR} obtained with the proposed framework with and without \ac{WBR}.}
\label{fig:improvements_of_WBR}
\end{figure}

Fig.~\ref{fig:improvements_of_WBR} illustrates the benefit of applying \ac{WBR} as a post-processing step to the initial channel estimates obtained using the methods in Section~\ref{initial_estimation}, prior to solving \eqref{eq:convex_problem}. \ac{WBR} consistently improves the \ac{MSE} across the considered estimators and \ac{SNR} range. At low \ac{SNR}, however, the gain is marginal because the most dominant spatial-frequency component in \eqref{eq:mu_identification} cannot be reliably identified due to the strong noise. As the \ac{SNR} increases, the dominant component is identified more reliably, resulting in more pronounced gains.

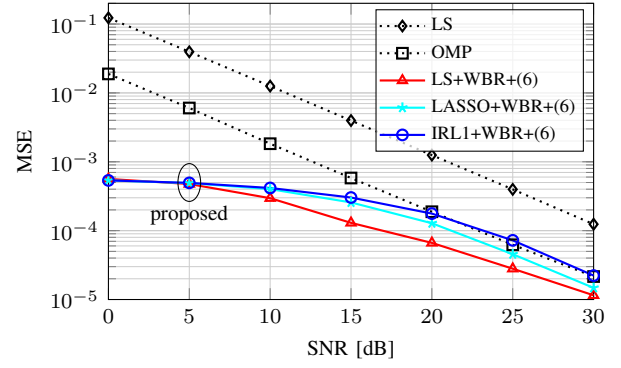
\begin{figure}
\centering
\input{Figures/fig_6}
\caption{\ac{MSE} versus \ac{SNR} obtained with the proposed framework compared with conventional estimators.}
\label{fig:perfomance_of_proposed}
\end{figure}

Fig.~\ref{fig:perfomance_of_proposed} compares the proposed framework with conventional \ac{LS} and \ac{OMP}~\cite{pati1993omp}, where the latter iteratively selects the most dominant wavenumber-domain channel components from the received pilot signal. For a fair comparison, the empirical analysis in Fig.~\ref{fig:analysis_C} determining the window length for \ac{WBR} is also used to set the \ac{OMP} sparsity level to $9$. \ac{OMP} and \ac{WBR} exploit the wavenumber-domain sparsity differently. As illustrated in Fig.~\ref{fig:effect_of_WBR}, the dominant channel power is localized around the strongest spatial-frequency component. \ac{WBR} leverages this structure by retaining a contiguous window around the dominant component, whereas \ac{OMP} selects components individually without enforcing any spatial-frequency contiguity, making it more susceptible to noisy pilot observations. Furthermore, the \ac{WBR}-refined channel estimate is subsequently enhanced by solving \eqref{eq:convex_problem}, which jointly exploits the sparsity and low-rank structure of the wavenumber-domain channel. As a result, the proposed framework significantly outperforms both benchmarks across the considered \ac{SNR} range, with the performance gap reducing at high \ac{SNR}.

\section{Conclusion}

In this paper, we presented a structured wavenumber-domain channel estimation framework that progressively refines an initial estimate by exploiting the localized spatial-frequency power distribution through \ac{WBR} and the sparsity and low-rank structure of the channel through a convex constrained formulation. The corresponding constraint bounds were selected using a conservative worst-case characterization over the relative \ac{Tx}--\ac{Rx} position. Numerical results showed that the proposed framework improves channel estimation accuracy, particularly at low \ac{SNR}. Future work will focus on deriving analytical methods for selecting the sparsity and rank constraint bounds and on developing a more principled design of the \ac{WBR} strategy.

\bibliographystyle{IEEEtran}
\bibliography{refs_abbr,di}

\end{document}

%% file: Acronyms.tex
\begin{acronym}
\acro{6G}{sixth-generation}
\acro{xlmimo}[XL-MIMO]{extremely large-scale multiple-input multiple-output}
\acro{mmWave}{millimeter wave}
\acro{sub-THz}{sub-terahert}
\acro{LoS}{line-of-sight}
\acro{NLoS}{non-line-of-sight}
\acro{CSI}{channel state information}
\acro{Tx}{transmitter}
\acro{Rx}{receiver}
\acro{ULA}{uniform linear array}
\acro{AWGN}{additive white Gaussian noise}
\acro{LS}{least-squares}
\acro{LASSO}{least absolute shrinkage and selection operator}
\acro{IRL1}{iteratively reweighted $\ell_1$-norm}
\acro{SNR}{signal-to-noise ratio}
\acro{MSE}{mean squared error}
\acro{NMSE}{normalized mean squared error}
\acro{OMP}{orthogonal matching pursuit}
\acro{WBR}{windowing-based refinement}
\acro{MIMO}{multiple-input multiple-output}
\acro{iid}[i.i.d.]{independent and identically distributed}
\end{acronym}

%% file: Definitions.tex
\newcommand{\Compl}{\mbox{$\mathbb{C}$}}

\newcommand{\argmax}{\operatornamewithlimits{argmax}}
\newcommand{\argmin}{\operatornamewithlimits{argmin}}

\newcommand{\Exp}{\mathbb{E}}

\renewcommand{\vec}{\mathrm{vec}}

\definecolor{oulu_blue}{HTML}{23408F}
\definecolor{oulu_green}{HTML}{39B54A}

\definecolor{red}{rgb}{1,0,0}
\definecolor{red_magenta}{rgb}{1,0,0.5}
\definecolor{magenta}{rgb}{1,0,1}
\definecolor{blue_magenta}{rgb}{0.5,0,1}
\definecolor{blue}{rgb}{0,0,1}
\definecolor{blue_cyan}{rgb}{0,0.5,1}
\definecolor{cyan}{rgb}{0,1,1}
\definecolor{green_cyan}{rgb}{0,1,0.5}
\definecolor{green}{rgb}{0,1,0}
\definecolor{green_yellow}{rgb}{0.5,1,0}
\definecolor{yellow}{rgb}{1,1,0}
\definecolor{red_yellow}{rgb}{1,0.5,0}

%% file: Figures/fig_1.tex
\begin{tikzpicture}[>=latex]
\small
\def\dh{1cm} 
\def\dv{1cm} 
\def \xcor{1cm}    
\def \ycor{-2cm}
\node[] (center) at (0,0) {};

\def \xsource {3.5};
\def \ysource {1.5};
\def \xpn {-2};
\def \rad {sqrt(\xsource^2+\ysource^2)};   
\def \ang {asin(\xsource/\rad)};            

\draw[->, thick, red] (\xcor,\ycor) -- (\xcor,\ycor+1cm)
    node[midway, sloped, above, xshift=2pt] {\footnotesize $x$-axis};
\draw[->, thick, red] (\xcor,\ycor) -- (\xcor+1cm,\ycor)
    node[midway, sloped, below, xshift=2pt] {\footnotesize $y$-axis};
\fill[red] (\xcor,\ycor) circle (2pt);

\draw[very thick] (-2.50*\dh,-2.5*\dv) -- (-2.50*\dh,1.5*\dv)
    node[sloped, above, yshift=2pt] {\footnotesize Rx};
\foreach \y in {-2.5,-2,...,1.5}{
    \node at (-2.50*\dh,\y*\dv) {$\times$};
}

\draw[very thick] (1.0*\dh, 0.0*\dv) -- (1.0*\dh, 2.0*\dv)
    node[sloped, above, yshift=2pt] {\footnotesize Tx};
\foreach \y in {0.0, 0.5, ..., 2.0}{
    \node at (1.0*\dh, \y*\dv) {$\times$};
}

\draw[thin] (-2.50*\dh,-0.5) -- (1.0*\dh, 1.0*\dv)
    node[midway, above, yshift=2pt] {\footnotesize $d$};

\draw[thin, domain=90-\ang:90]
    plot ({1*cos(\x)-2.5}, {1*sin(\x)-0.5});
\node[anchor=west] at (-2.3, 0) {$\theta$};

\draw[thick, dashed, red_magenta, domain=160:240]
    plot ({2.2*cos(\x)+1}, {2.2*sin(\x)+1});
\draw[thick, dashed, red_magenta, domain=160:240]
    plot ({3.808*cos(\x)+1}, {3.808*sin(\x)+1});

\draw[<->, black] (-3*\dh, -2.5*\dv) -- (-3*\dh,  1.5*\dv)
    node[midway, anchor=east]  {$L_{\textrm{r}}$};
\draw[<->, black] ( 1.5*\dh,  0.0*\dv) -- ( 1.5*\dh,  2.0*\dv)
    node[midway, anchor=west]  {$L_{\textrm{t}}$};        
\draw[<->, black] (-2.25*\dh,  1.0*\dv) -- (-2.25*\dh,  1.5*\dv)
    node[midway, anchor=west]  {$\delta_{\textrm{r}}$};
\draw[<->, black] ( 0.75*\dh,  1.5*\dv) -- ( 0.75*\dh,  2.0*\dv)
    node[midway, anchor=east]  {$\delta_{\textrm{t}}$};   
\end{tikzpicture}

%% file: Figures/fig_2.tex
\begin{tikzpicture}

\begin{axis}[
	width=4.8cm, height=4.5cm,
	xmin=1, xmax=8,
	ymin=0, ymax=40,
	xlabel={$\sqrt{N_\textrm{t}}$},
	ylabel={$\alpha$},
	xlabel near ticks,
	ylabel near ticks,
	label style={font=\footnotesize},
	xtick={2,4,...,8},
	ytick={0,10,...,40},
    ticklabel style={font=\footnotesize},
	legend style={at={(0.98,0.02)}, anchor=south east, font=\scriptsize, inner sep=1pt, fill opacity=0.75, draw opacity=1, text opacity=1},
	legend cell align=left,
	title style={font=\scriptsize, yshift=-2mm},
	grid=major,
    grid style={line width=.1pt, draw=gray!40},
]

\addplot[thick, black, mark=o]
table [x=sqrt_Nt, y=alpha_true, col sep=comma] {Figures/data.txt/fig_2_data/alpha_Nr64.txt};
\addlegendentry{True};

\addplot[thick, black, dashed, mark=asterisk, mark options=solid]
table [x=sqrt_Nt, y=alpha_linear, col sep=comma] {Figures/data.txt/fig_2_data/alpha_Nr64.txt};
\addlegendentry{Linear};

\addplot[thick, red, mark=o]
table [x=sqrt_Nt, y=alpha_true, col sep=comma] {Figures/data.txt/fig_2_data/alpha_Nr128.txt};

\addplot[thick, red, dashed, mark=asterisk, mark options=solid]
table [x=sqrt_Nt, y=alpha_linear, col sep=comma] {Figures/data.txt/fig_2_data/alpha_Nr128.txt};

\addplot[thick, blue, mark=o]
table [x=sqrt_Nt, y=alpha_true, col sep=comma] {Figures/data.txt/fig_2_data/alpha_Nr256.txt};

\addplot[thick, blue, dashed, mark=asterisk, mark options=solid]
table [x=sqrt_Nt, y=alpha_linear, col sep=comma] {Figures/data.txt/fig_2_data/alpha_Nr256.txt};

\begin{scope}[>=latex]
\draw[->] (4,33) -- (5.5,15) {};
\end{scope}

\node[black, font=\scriptsize, anchor=south] at (4,32) {$N_{\textrm{r}} = 256, 128, 64$};

\end{axis}
\end{tikzpicture} \hspace{-3mm}
\begin{tikzpicture}
\begin{axis}[
	width=4.8cm, height=4.5cm,
	xmin=1, xmax=64,
	ymin=1, ymax=2,
	xlabel={$N_\textrm{t}$},
	ylabel={$\beta$},
	xlabel near ticks,
	ylabel near ticks,
	label style={font=\footnotesize},
	xtick={4,8,16,32,64},
	ytick={1,1.2,...,2},
    ticklabel style={font=\footnotesize},
	legend style={at={(0.02,0.98)},anchor=north west, font=\scriptsize, inner sep=1pt, fill opacity=0.75, draw opacity=1, text opacity=1},
	legend cell align=left,
	title style={font=\scriptsize, yshift=-2mm},
	grid=major,
    grid style={line width=.1pt, draw=gray!40},
]

\addplot[thick, black, mark=o]
table [x=Nt, y=beta_true, col sep=comma] {Figures/data.txt/fig_2_data/beta_Nr64.txt};
\addlegendentry{True};

\addplot[thick, black, dashed, mark=asterisk, mark options=solid]
table [x=Nt, y=beta_linear, col sep=comma] {Figures/data.txt/fig_2_data/beta_Nr64.txt};
\addlegendentry{Linear};

\addplot[thick, red, mark=o]
table [x=Nt, y=beta_true, col sep=comma] {Figures/data.txt/fig_2_data/beta_Nr128.txt};

\addplot[thick, red, dashed, mark=asterisk, mark options=solid]
table [x=Nt, y=beta_linear, col sep=comma] {Figures/data.txt/fig_2_data/beta_Nr128.txt};

\addplot[thick, blue, mark=o]
table [x=Nt, y=beta_true, col sep=comma] {Figures/data.txt/fig_2_data/beta_Nr256.txt};

\addplot[thick, blue, dashed, mark=asterisk, mark options=solid]
table [x=Nt, y=beta_linear, col sep=comma] {Figures/data.txt/fig_2_data/beta_Nr256.txt};

\begin{scope}[>=latex]
\draw[->] (42,1.1) -- (26,1.6) {};
\end{scope}

\node[black, font=\scriptsize, anchor=south] at (42,0.97) {$N_{\textrm{r}} = 256, 128, 64$};

\end{axis}
\end{tikzpicture} 

%% file: Figures/fig_3.tex
\begin{tikzpicture}

\begin{axis}[
	width=3.2cm, height=3.8cm,
	xmin=-55, xmax=-20,
	ymin=0, ymax=1.1,
	xlabel={$\zeta$},
	ylabel={$\big|\mathbf{H}_\textrm{w}(:,i)\big|$},
	xlabel near ticks,
	ylabel near ticks,
	label style={font=\footnotesize},
	xtick={-55,-40,...,-20},
	ytick={0,0.2,...,1},
    ticklabel style={font=\footnotesize},
	legend style={at={(0.98,0.98)}, anchor=north east, font=\scriptsize, inner sep=1pt, fill opacity=0.75, draw opacity=1, text opacity=1},
	legend cell align=left,
	title style={font=\scriptsize, yshift=-2mm},
	grid=major,
    grid style={line width=.1pt, draw=gray!40},
]

\addplot[ybar,bar width=1pt, draw=black, fill=blue, ybar legend]
table [x=Var1, y=Var2, col sep=comma] {Figures/data.txt/fig_3_data/true.csv};

\end{axis}
\end{tikzpicture} \hspace{-3mm}
\begin{tikzpicture}

\begin{axis}[
	width=3.2cm, height=3.8cm,
	xmin=-55, xmax=-20,
	ymin=0, ymax=1.1,
	xlabel={$\zeta$},
	ylabel={$\big|\tilde{\mathbf{H}}_\textrm{w}(:,i)\big|$},
	xlabel near ticks,
	ylabel near ticks,
	label style={font=\footnotesize},
	xtick={-55,-40,...,-20},
	ytick={0,0.2,...,1},
    ticklabel style={font=\footnotesize},
	legend style={at={(0.5,1.15)}, anchor=north, font=\scriptsize, inner sep=1pt, fill opacity=0.5, draw opacity=1, text opacity=1},
	legend cell align=left,
	title style={font=\scriptsize, yshift=-2mm},
	grid=major,
    grid style={line width=.1pt, draw=gray!40},
]

\addplot[thick , red]
table [x=Var1, y=windowfn, col sep=comma] {Figures/data.txt/fig_3_data/WindowFn.csv};
\addlegendentry{$\boldsymbol{\psi}_i(C)$};

\addplot[ybar, bar width=1pt, draw=black, fill=blue]
table [x=Var1, y=Var2, col sep=comma] {Figures/data.txt/fig_3_data/noisy_est.csv};

\end{axis}
\end{tikzpicture} \hspace{-3mm}
\begin{tikzpicture}

\begin{axis}[
	width=3.2cm, height=3.8cm,
	xmin=-55, xmax=-20,
	ymin=0, ymax=1.1,
	xlabel={$\zeta$},
	ylabel={$\big|\tilde{\mathbf{H}}_\textrm{w,WBR}(:,i)\big|$},
	xlabel near ticks,
	ylabel near ticks,
	label style={font=\footnotesize},
	xtick={-55,-40,...,-20},
	ytick={0,0.2,...,1},
    ticklabel style={font=\footnotesize},
	legend style={at={(0.98,0.98)}, anchor=north east, font=\scriptsize, inner sep=1pt, fill opacity=0.75, draw opacity=1, text opacity=1},
	legend cell align=left,
	title style={font=\scriptsize, yshift=-2mm},
	grid=major,
    grid style={line width=.1pt, draw=gray!40},
]

\addplot[ybar, bar width=1pt, draw=black, fill=blue, ybar legend]
table [x=Var1, y=Var2, col sep=comma] {Figures/data.txt/fig_3_data/Windowed_est.csv};

\end{axis}
\end{tikzpicture}

%% file: Figures/fig_4.tex
\pgfdeclarelayer{background}
\pgfdeclarelayer{foreground}
\pgfsetlayers{background,main,foreground}

\begin{tikzpicture}

\begin{pgfonlayer}{background}
\begin{axis}[
	name=main3d,
	width=6cm,
	height=5.5cm,
	xmin=0, xmax=90,
	ymin=20, ymax=80,
	zmin=0, zmax=10,
	view={-30}{30},
    colormap name = hot,
	xlabel={$\theta$ [deg]},
	ylabel={$d$ [m]},
	zlabel={$N_\text{eff}$},
	x label style={at={(axis cs:50,0,1e-3)}, anchor=north},
	y label style={at={(axis cs:-20,50,-1)}, anchor=north},
	label style={font=\footnotesize},
	ticklabel style={font=\footnotesize},
	xtick={0,15,...,90},
	ytick={20,40,...,80},
	ztick={1,3,...,10},
	grid=major,
    grid style={line width=.1pt, draw=gray!40},
	mesh/rows=500,
	mesh/cols=10,
]

\addplot3[
	surf,
	opacity=0.90
]
table[
	x=theta,
	y=r,
	z=eff_sf,
	col sep=comma
]{Figures/data.txt/fig_4_data/plotdata.csv};

\end{axis}
\end{pgfonlayer}

\hspace{2mm}

\begin{pgfonlayer}{foreground}
\begin{axis}[
	axis background/.style={fill=white},
	at={(main3d.outer north east)},
	anchor=outer north west,
	width=4.3cm,
	height=2.5cm,
	xmin=0, xmax=60,
	ymin=0, ymax=10,
	xlabel={$\theta$ [deg]},
	label style={font=\tiny},
	ticklabel style={font=\tiny},
	xtick={0,10,...,90},
	ytick={0,5,...,10},
	title={$d \approx 80$\,m},
	title style={font=\scriptsize, yshift=-1mm},
	grid=both,
    grid style={line width=.1pt, draw=gray!40},
]

\addplot[thin, black, mark=diamond,mark size=1.2pt,mark repeat = 5] table[x=Var1, y=Var3, col sep=comma]{Figures/data.txt/fig_4_data/80m_plotdata.csv}; 

\end{axis}
\end{pgfonlayer} \vspace{2 cm}

\begin{pgfonlayer}{foreground}
\begin{axis}[
	axis background/.style={fill=white},
	at={(main3d.outer south east)},
	anchor=outer south west,
	width=4.3cm,
	height=2.5cm,
	xmin=0, xmax=60,
	ymin=0, ymax=10,
	xlabel={$\theta$ [deg]},
	label style={font=\tiny},
	ticklabel style={font=\tiny},
	xtick={0,10,...,90},
	ytick={0,5,...,10},
	title={$d \approx 20$\,m},
	title style={font=\scriptsize, yshift=-1mm},
	grid=both,
	grid style={line width=.1pt, draw=gray!40},
]

\addplot[thin, black, mark=diamond,mark size=1.2pt,mark repeat = 5] table[x=Var1, y=Var3, col sep=comma]{Figures/data.txt/fig_4_data/20m_plotdata.csv}; 

\end{axis}
\end{pgfonlayer}

\end{tikzpicture}

%% file: Figures/fig_5.tex
\pgfdeclarelayer{background}
\pgfdeclarelayer{foreground}
\pgfsetlayers{background,main,foreground}

\begin{tikzpicture}
\begin{pgfonlayer}{background}
\begin{axis}[
    width=8cm,
    height=5.5cm,
    ymode=log,
    xmin=0, xmax=30,
    ymin=1e-5, ymax=7e-4,
    xtick={0,5,...,30},
    ytick={1e-5, 1e-4,1e-3,1e-2,1e-1,1},
    ticklabel style={font=\footnotesize},
    xlabel={SNR [dB]},
    ylabel={MSE},
    xlabel near ticks,
	ylabel near ticks,
	label style={font=\footnotesize},
    legend style={at={(0.98,0.98)}, anchor=north east, font=\scriptsize, inner sep=1pt, fill opacity=0.75, draw opacity=1, text opacity=1},
	legend cell align=left,
    grid=both,
    grid style={line width=.1pt, draw=gray!40},
]

\addplot[thick, black]
table [x=Var1, y=Var2, col sep=comma] {Figures/data.txt/fig_5_data/L1_optimized_with_fil.csv};
\addlegendentry{With WBR};

\addplot[thick, black, dashed]
table [x=Var1, y=Var2, col sep=comma] {Figures/data.txt/fig_5_data/L1_optimized_without_fil.csv};
\addlegendentry{Without WBR};

\addplot[thick, blue, dashed, mark options=solid, mark=o]
table [x=Var1, y=Var2, col sep=comma] {Figures/data.txt/fig_5_data/L1_optimized_without_fil.csv};
\addplot[thick, blue, mark=o]
table [x=Var1, y=Var2, col sep=comma] {Figures/data.txt/fig_5_data/L1_optimized_with_fil.csv};

\addplot[thick, cyan, dashed, mark options=solid, mark=star]
table [x=Var1, y=Var2, col sep=comma] {Figures/data.txt/fig_5_data/Lasso_optimized_without_fil.csv};
\addplot[thick, cyan, mark=star]
table [x=Var1, y=Var2, col sep=comma] {Figures/data.txt/fig_5_data/Lasso_optimized_with_fil.csv};

\addplot[thick, red, dashed, mark options=solid, mark=triangle]
table [x=Var1, y=Var2, col sep=comma] {Figures/data.txt/fig_5_data/LS_optimized_without_fil.csv};
\addplot[thick, red, mark=triangle]
table [x=Var1, y=Var2, col sep=comma] {Figures/data.txt/fig_5_data/LS_optimized_with_fil.csv};

\coordinate (zoom_coord) at (15,9.7656e-05);
\draw[black] (4,5.3711e-04) rectangle (6,4.3945e-04) {};

\coordinate (mainLS) at (axis cs:8,1.7e-4);
\coordinate (mainLASSO) at (axis cs:17,3.5e-5);
\coordinate (mainL1) at (axis cs:20,1.5e-5);

\node[black, font=\scriptsize, anchor=west] at (17,3.5e-5) {$\textrm{LASSO}$};
\node[black, font=\scriptsize, anchor=west] at (7,2e-4) {$\textrm{LS}$};
\node[black, font=\scriptsize, anchor=west] at (20,1.5e-5) {$\textrm{IRL1}$};

\begin{scope}[>=latex]
\draw[->] (axis cs:9,2.4e-4) -- (axis cs:10,3e-4) {};  
\draw[->] (axis cs:21.5,4e-5) -- (axis cs:23.8,6e-5) {}; 
\draw[->] (axis cs:23,1.8e-5) -- (axis cs:28,6e-5) {}; 

\draw[->] (axis cs:5,4.3945e-04) -- (axis cs:5,1e-4) {};
\end{scope}

\end{axis}
\end{pgfonlayer}

\begin{pgfonlayer}{foreground}
\begin{axis}[
	axis background/.style={fill=white},
	at={(zoom_coord)},
	anchor={outer north east},
	width=0.25\textwidth,
	height=0.18\textwidth,
	xmin=4, xmax=6,
	ymin=4.3945e-04, ymax=5.3711e-04,
	ticks=none,
]

\addplot[thick, blue, dashed, mark options=solid, mark=o]
table [x=Var1, y=Var2, col sep=comma] {Figures/data.txt/fig_5_data/L1_optimized_without_fil.csv};
\addplot[thick, blue, mark=o]
table [x=Var1, y=Var2, col sep=comma] {Figures/data.txt/fig_5_data/L1_optimized_with_fil.csv};

\addplot[thick, cyan, dashed, mark options=solid, mark=star]
table [x=Var1, y=Var2, col sep=comma] {Figures/data.txt/fig_5_data/Lasso_optimized_without_fil.csv};
\addplot[thick, cyan, mark=star]
table [x=Var1, y=Var2, col sep=comma] {Figures/data.txt/fig_5_data/Lasso_optimized_with_fil.csv};

\addplot[thick, red, dashed, mark options=solid, mark=triangle]
table [x=Var1, y=Var2, col sep=comma] {Figures/data.txt/fig_5_data/LS_optimized_without_fil.csv};
\addplot[thick, red, mark=triangle]
table [x=Var1, y=Var2, col sep=comma] {Figures/data.txt/fig_5_data/LS_optimized_with_fil.csv};

\coordinate (zoomLS) at (axis cs:5,5.2e-04);
\coordinate (zoomLASSO) at (axis cs:5.2,5e-4);
\coordinate (zoomL1) at (axis cs:5.8,4.8e-04);

\end{axis}
\begin{scope}[>=latex]
\draw[->] (mainLS) -- (zoomLS);
\draw[->] (mainLASSO) -- (zoomLASSO);
\draw[->] (mainL1) -- (zoomL1);
\end{scope}
\end{pgfonlayer}

\end{tikzpicture}

%% file: Figures/fig_6.tex
\begin{tikzpicture}

\begin{axis}[
    width=8cm,
    height=5.5cm,
    ymode=log,
    xmin=0, xmax=30,
    ymin=1e-5, ymax=2e-1,
    xtick={0,5,...,30},
    ytick={1e-5, 1e-4,1e-3,1e-2,1e-1,1,10,100,1000},
    ticklabel style={font=\footnotesize},
    xlabel={SNR [dB]},
    ylabel={MSE},
    xlabel near ticks,
	ylabel near ticks,
	label style={font=\footnotesize},
    legend style={at={(0.98,0.98)}, anchor=north east, font=\scriptsize, inner sep=1pt, fill opacity=0.75, draw opacity=1, text opacity=1},
	legend cell align=left,
    grid=both,
    grid style={line width=.1pt, draw=gray!40},
]

\addplot[thick, black, dotted, mark options=solid, mark=diamond]
table [x=Var1, y=Var2, col sep=comma] {Figures/data.txt/fig_6_data/LS_ori.csv};
\addlegendentry{LS};

\addplot[thick, black, dotted, mark options=solid, mark=square]
table [x=Var1, y=Var2, col sep=comma] {Figures/data.txt/fig_6_data/OMP_ori.csv};
\addlegendentry{OMP};

\addplot[thick, red, mark=triangle]
table [x=Var1, y=Var2, col sep=comma] {Figures/data.txt/fig_6_data/LS_optimized_with_fil.csv};
\addlegendentry{LS+WBR+\eqref{eq:convex_problem}};

\addplot[thick, cyan, mark=star]
table [x=Var1, y=Var2, col sep=comma] {Figures/data.txt/fig_6_data/Lasso_optimized_with_fil.csv};
\addlegendentry{LASSO+WBR+\eqref{eq:convex_problem}};

\addplot[thick, blue, mark=o]
table [x=Var1, y=Var2, col sep=comma] {Figures/data.txt/fig_6_data/L1_optimized_with_fil.csv};
\addlegendentry{IRL1+WBR+\eqref{eq:convex_problem}};

\draw[black, thin](axis cs:5,5e-4) ellipse [x radius=1.5mm, y radius=2.5mm];
\node[anchor=south] at (5,9e-5) {\footnotesize{proposed}};

\end{axis}
\end{tikzpicture}